\documentclass[
 reprint,preprintnumbers,
 amsmath,amssymb,
 aps,showpacs
]{revtex4-1}
\usepackage{bm}
\usepackage{subfigure}
\usepackage[dvipdfmx]{graphicx}
\usepackage{color}

\allowdisplaybreaks[1]

\begin{document}

\title{Photon production spectrum above $T_{\rm c}$ with a lattice quark propagator}
\author{Taekwang Kim}
 \email{kim@kern.phys.sci.osaka-u.ac.jp}
\affiliation{Department of Physics, Osaka University,
Toyonaka, Osaka 560-0043, Japan}
\author{Masayuki Asakawa}
 \email{yuki@phys.sci.osaka-u.ac.jp}
\affiliation{Department of Physics, Osaka University,
Toyonaka, Osaka 560-0043, Japan}
\author{Masakiyo Kitazawa}
 \email{kitazawa@phys.sci.osaka-u.ac.jp}
\affiliation{Department of Physics, Osaka University,
Toyonaka, Osaka 560-0043, Japan}
\affiliation{
J-PARC Branch, KEK Theory Center,
Institute of Particle and Nuclear Studies, KEK,
203-1, Shirakata, Tokai, Ibaraki, 319-1106, Japan }
\date{\today}

\begin{abstract}
The photon production rate from the deconfined medium is analyzed 
with the photon self-energy constructed from the quark propagator 
obtained by the numerical simulation on the quenched lattice
for two values of temperature, $T=1.5T_{\rm c}$ and $3T_{\rm c}$,
above the critical temperature $T_{\rm c}$. 
The photon self-energy is calculated by the Schwinger-Dyson 
equation with the lattice quark propagator and a vertex function 
determined so as to satisfy the Ward-Takahashi identity.
The obtained photon production rate 
exhibits a similar behavior as the perturbative results
at the energy of photons larger than $0.5$~GeV.
\end{abstract}

\preprint{J-PARC-TH-0096}

\maketitle 

\def\slash#1{\not\hspace{-1mm}#1}
\def\slashb#1{\not\!\!#1}
\def\delsla{\not\!\partial}

\section{Introduction}
\label{sec:Introduction}

Photons are important observables in relativistic heavy-ion 
collisions~\cite{FIRST,Muller:2012zq}.
Because of its colorless nature, the mean free path of 
photons is longer than the size of the hot matter produced in
heavy-ion collisions.
Once photons are produced, therefore, they penetrate
the hot medium without scatterings via the strong interaction.
The experimental measurements of photon yield and anisotropic flow have 
been carried out actively~\cite{Adare:2014fwh,Adam:2015lda}.

The experimental results show that the photon yield from heavy-ion 
collisions has an excess at lower transverse momentum
regions~\cite{Adare:2014fwh,Adam:2015lda}
compared with the reference yield obtained by the superposition of 
proton-proton collisions.
This result suggests the importance of the photon radiation from 
the hot medium.
The other important observable is the photon anisotropic flow.
The magnitude of the observed photon flows is almost the same as that 
of the medium~\cite{Adare:2015lcd}.
The large anisotropic flow implies that photons are dominantly emitted 
from late stages in which the flow of the medium has been well developed.
Phenomenological analyses of photon yield based on dynamical models 
have not succeeded in reproducing these experimental results simultaneously
so far~\cite{Paquet:2015lta,Shen:2013cca}.
Photons are produced via several mechanisms
during the time evolution of heavy-ion collisions.
The discrepancy between the phenomenological analysis and experimental 
results suggests the importance of careful surveys of the production 
rates through these different production mechanisms.
In the present study, among these mechanisms
we focus on the photon radiation from the hot deconfined medium.

In the previous phenomenological analyses on the photon 
yield and flows~\cite{Paquet:2015lta,Shen:2013cca}, 
the photon production rate calculated in perturbation 
theory~\cite{Kapusta:1991qp,Baier:1991em,Arnold:2001ba,Arnold:2001ms,
Ghiglieri:2013gia} has been used for those from deconfined medium. 
On the other hand, the success of hydrodynamic models in describing 
the time evolution of the hot medium suggests that the deconfined medium 
is a strongly coupled system~\cite{Hirano:2005wx,Hirano:2005xf}, 
in which a perturbative analysis would lose its validity.
A non-perturbative analysis thus is desirable to
evaluate the production rate from the deconfined medium.
However, the direct analysis of the emission rate in lattice
QCD numerical simulations is quite difficult, 
because this analysis requires an analysis of on-shell spectral 
functions.

In this paper, we analyze the photon production rate
from the deconfined medium by partially using the lattice data
in order to incorporate the non-perturbative effects in the analysis.
The photon production rate is related to the photon self-energy
in medium~\cite{McLerran:1984ay,Weldon:1990iw,Gale:1990pn}.
From the Schwinger-Dyson equation~(SDE), the full photon self-energy
is constructed from the
full quark propagator and the full photon-quark vertex.
In Refs.~\cite{Karsch:2007wc,Karsch:2009tp,Kaczmarek:2012mb}, 
the quark propagator in the deconfined medium is analyzed on the 
quenched lattice in Landau gauge with a pole ansatz.
We employ this lattice propagator as the full quark propagator 
in the SDE.
A photon-quark vertex is constructed from the lattice quark propagator
so as to satisfy the Ward-Takahashi identity.
Our analysis thus respects gauge invariance.
The same analysis has been applied to the calculation of the dilepton 
production rate at zero three-momentum in our previous study~\cite{Kim:2015poa}.
The present study extends this analysis to nonzero momentum to calculate
the photon production rate.

We analyze the photon production rate for two temperatures,
$T=1.5T_{\rm c}$ and $3T_{\rm c}$, where
$T_{\rm c}$ is the pseudo critical temperature.
We show that the photon production is dominated by the pair annihilation
and Landau damping of two quasi-particle, the normal and plasmino,
modes.
Although the pair annihilation cannot produce real photons in the perturbative 
analysis, in our formalism peculiar dispersion relations of quasi-particle 
modes in the lattice quark propagator allow the process.
We find that the obtained photon production rate exhibits a similar behavior 
as the perturbative results both in the slope and the magnitude 
at the photon energy larger than $0.5$~GeV, 
although the photon production mechanism is completely different.

The outline of this paper is as follows.
In the next section,
we introduce the SDE for the photon self-energy, and 
discuss the construction of its components, the lattice quark propagator 
and a photon-quark vertex satisfying the Ward-Takahashi identity.
In Sec.~\ref{sec:PPR},
we calculate the photon self-energy with this setup 
with and without vertex correction.
In Sec.~\ref{sec:NR}, we show the numerical results on the 
photon production spectrum.
The final section is devoted to a summary.

\section{Schwinger-Dyson equation for Photon Self-Energy}
\label{sec:SDEfPSE}

\subsection{Schwinger-Dyson equation}
\label{sec:SDE}

\begin{figure}
\begin{center}
\includegraphics[width=0.35\textwidth]{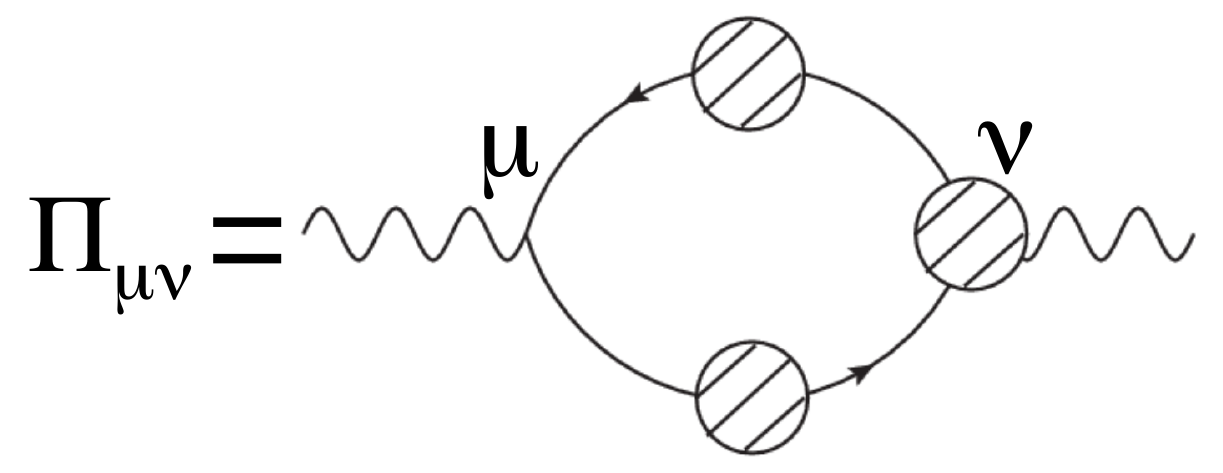}
\end{center}
\caption{
Diagrammatic representation of the Schwinger-Dyson equation for 
the full photon self-energy, Eq.~(\ref{eq:SDE_Pi_Start}). 
The shaded circles mean the full quark propagator and 
the full photon-quark vertex.}
\label{fig:SDE_PSE}
\end{figure}

The photon production rate from a static medium per unit time per unit volume
is related to the retarded photon self-energy $\Pi_{\mu\nu}^R(\omega,\bm{q})$
as~\cite{McLerran:1984ay,Weldon:1990iw,Gale:1990pn}
\begin{align}
  \omega\frac{d N_\gamma}{d^3qd^4x}
  = -\frac{1}{(2\pi)^3}\frac{1}{{\rm e}^{\beta\omega}-1}{\rm Im}
  \Pi_\mu^{R,\mu}(\omega,\bm{q}),
  \label{eq:SDE_PhotonRate}
\end{align}
with the inverse temperature $\beta=1/T$
and the four-momentum of
photon~$(\omega,\bm{q})$.
This formula is correct at the leading order in
electromagnetic interaction and all orders in strong interaction.
The retarded photon self-energy is related to 
the imaginary-time self-energy $\Pi_{\mu\nu}(i\omega_m,\bm{q})$
by the analytic continuation,
\begin{align}
\Pi^R_{\mu\nu}(\omega,\bm{q}) = 
\Pi_{\mu\nu}(i\omega_m,\bm{q})|_{i\omega_m \rightarrow \omega+i\eta},
\label{eq:SDE_AnalyticCon}
\end{align}
with the positive infinitesimal number $\eta$
and the Matsubara frequency for bosons~$\omega_m=2\pi Tm$ 
with integer $m$.
The imaginary-time full photon self-energy is given by 
the SDE in Matsubara formalism with the full quark propagator~$S(P)$
and the full photon-quark vertex~$\Gamma_\mu(P+Q,P)$ as
\begin{align}
\Pi_{\mu\nu}(i\omega_m,\bm{q})=&-\sum_fe_f^2T\sum_n
\int\frac{d^3p}{(2\pi)^3}
\nonumber
\\
&{\rm Tr_C Tr_D}[S(P)\gamma_\mu S(P+Q)\Gamma_\nu(P+Q,P)],
\label{eq:SDE_Pi_Start}
\end{align}
where
$P_\mu=(i\nu_n,\bm{p})$ and $Q_\mu=(i\omega_m,\bm{q})$
are the four-momenta of quarks and photons, respectively, 
and $\nu_n=(2n+1)\pi T$ with integer $n$ is the Matsubara frequency 
for fermions.
$e_f$ is the electric charge of quarks with the
index $f$ representing the quark flavor.
The color, flavor, and Dirac indices of $S(P)$ are
suppressed for notational simplicity, and
${\rm Tr_C}$ and ${\rm Tr_D}$ represent
the trace over the color and Dirac indices, respectively.
In covariant gauges,
the off-diagonal elements of the quark propagator in color space vanish.
As a result, the trace over the color indices gives a factor three
in Eq.~(\ref{eq:SDE_Pi_Start}).
In Fig.~\ref{fig:SDE_PSE},
we show the diagrammatic representation of Eq.~(\ref{eq:SDE_Pi_Start}).
The shaded circles mean the full propagator
and vertex function.
In the following, we consider the two-flavor QCD
with degenerate $u$ and $d$ quarks, in which
$\sum_fe_f^2=5e^2/9$.

\subsection{Lattice quark propagator and spectral function}
\label{sec:LQPaSF}

In the present study,
we employ the quark propagator obtained in the lattice numerical simulation 
in Ref.~\cite{Karsch:2007wc,Karsch:2009tp,Kaczmarek:2012mb}
for the full quark propagator in Eq.~(\ref{eq:SDE_Pi_Start}).
This approach was used in our previous study for the analysis of 
the dilepton production rate at zero three-momentum~\cite{Kim:2015poa}.
In this subsection, we give a brief review on the
general property of the quark propagator and
discuss the implementation of the results in Ref.~\cite{Kaczmarek:2012mb}
in our analysis.

The imaginary-time quark propagator for momentum $\bm{p}$ is defined by
\begin{align}
S_{\mu\nu}(\tau,\bm{p})=\int d^3x{\rm e}^{-i\bm{p}\cdot\bm{x}}
\langle  \psi_\mu(\tau,\bm{x})\bar{\psi}_\nu(0,\bm{0})\rangle,
\label{eq:LQPaSF_S(tau)}
\end{align}
where $\psi_\mu(\tau,\bm{x})$ is the quark field with the
Dirac index $\mu$, and $\tau$ is the imaginary time
restricted to the interval $0\leq \tau < \beta$.
We show the Dirac indices of quark fields
explicitly for a moment.
The Fourier transform of Eq.~(\ref{eq:LQPaSF_S(tau)}) is given by
\begin{align}
  S_{\mu\nu}(i\nu_n,\bm{p})
  = \int_0^\beta d\tau  
  {\rm e}^{i\nu_n\tau} S_{\mu\nu}(\tau,\bm{p}).
\label{eq:LQPaSF_S(nu)}
\end{align}
Equation~(\ref{eq:LQPaSF_S(nu)}) is rewritten with the
quark spectral function $\rho_{\mu\nu}(\nu,\bm{p})$ as
\begin{align}
  S_{\mu\nu}(i\nu_n,\bm{p}) 
  = -\int_{-\infty}^\infty d\nu'\frac{\rho_{\mu\nu}(\nu',\bm{p})}{\nu'-i\nu_n}.
  \label{eq:LQPaSF_S}
\end{align}
From Eqs.~(\ref{eq:LQPaSF_S(nu)}) and (\ref{eq:LQPaSF_S}), 
the spectral function is related to Eq.~(\ref{eq:LQPaSF_S(tau)}) as
\begin{align}
S_{\mu\nu}\left(\tau,\bm{p}\right) =
\int_{-\infty}^{\infty}d\nu
\frac{\mbox{e}^{\left(1/2-\tau/\beta\right)\beta \nu}}{\mbox{e}^{\beta \nu/2}+\mbox{e}^{-\beta \nu/2}}
\rho_{\mu\nu}\left(\nu,\bm{p}\right).
\label{eq:LQPaSF_S:rho}
\end{align}

When chiral symmetry is restored,
the quark propagator anticommutes with $\gamma_5$.
In this case, one can decompose the quark spectral function
with the projection operators
$\Lambda_{\pm}\left(\bm{p}\right)=\left(1\pm\gamma_0\hat{\bm{p}}\cdot\bm{\gamma}\right)/2$
as
\begin{align}
  \rho(\nu,\bm{p}) 
  = \rho_+(\nu,p) \Lambda_+(\bm{p}) \gamma_0 
  + \rho_-(\nu,p) \Lambda_-(\bm{p}) \gamma_0,
  \label{eq:LQPaSF_Decompose}
\end{align}
with $p=|\bm{p}|$, $\hat{\bm{p}}=\bm{p}/p$, and 
\begin{align}
  \rho_\pm(\nu,p) 
  = \frac12 \mbox{Tr}_{\rm D}
  [\rho(\nu,\bm{p})\gamma_0\Lambda_\pm(\bm{p})].
\end{align}
From the anticommutation relation of $\psi$ and $\bar{\psi}$,
one can derive the sum rule
\begin{align}
  \int_{-\infty}^\infty d\nu\rho_\pm(\nu,p)=1.
  \label{eq:LQPaSF_SumRule}
\end{align}
Using the charge conjugation symmetry of thermal ensemble, 
one can also show that $\rho_\pm$ satisfy~\cite{Kaczmarek:2012mb}
\begin{align}
  \rho_\pm(\nu,p)=\rho_\mp(-\nu,p).
  \label{eq:rho_pm_mp}
\end{align}

In Refs.~\cite{Karsch:2007wc,Karsch:2009tp,Kaczmarek:2012mb},
the imaginary-time quark propagator Eq.~(\ref{eq:LQPaSF_S(tau)}) is 
evaluated on the lattice in Landau gauge with
the quenched approximation.
On the lattice the quark propagator is measured for discrete imaginary 
times.
To obtain the real time information,
one has to deduce the spectral function from the discrete information.
In Refs.~\cite{Karsch:2007wc,Karsch:2009tp,Kaczmarek:2012mb},
the spectral function $\rho_\pm(\nu,p)$ is analyzed with the two-pole 
ansatz,
\begin{equation}
  \rho_+(\nu,p)
  = Z_+(p) \delta\left(\nu-\nu_+(p)\right)
  + Z_-(p) \delta\left(\nu+\nu_-(p)\right),
  \label{eq:LQPaSF_2pole}
\end{equation}
where $Z_\pm(p)$ and $\nu_\pm(p)$ are the residues and 
positions of the poles, respectively, which are determined by fitting
to the lattice propagator for each $p$.

Comments on the two-pole ansatz Eq.~(\ref{eq:LQPaSF_2pole}) are in order.
First, two particle-like states $\nu_+(p)$ and $\nu_-(p)$ in 
Eq.~(\ref{eq:LQPaSF_2pole}) correspond to 
the normal and plasmino modes, respectively, which appear in the quark 
propagator in the hard thermal loop approximation \cite{LeBellac}.
In fact, the analyses in 
Refs.~\cite{Karsch:2007wc,Karsch:2009tp,Kaczmarek:2012mb} suggest
that these modes behave like the normal and plasmino modes as functions 
of momentum $p$ and bare quark mass $m_0$.
For example, for large $p$ and $m_0$ the residue of the plasmino mode
$Z_-(p)$ becomes small and the propagator
approaches the one of the free quark.
Second, the restoration of chiral symmetry for massless quarks 
above the pseudo-critical temperature~$T_{\rm c}$ is confirmed
explicitly on the lattice by measuring the scalar term
of the quark propagator~\cite{Karsch:2009tp,Kaczmarek:2012mb}.
Third, in Refs.~\cite{Karsch:2009tp,Kaczmarek:2012mb}
various fitting ans\"atze have been tested besides the two-pole ansatz
Eq.~(\ref{eq:LQPaSF_2pole}), such as 
an ansatz which allows non-zero widths of the poles in
Eq.~(\ref{eq:LQPaSF_2pole}).
However, it was turned out that fitting with widths
does not improve $\chi^2$, as the minimum of the $\chi^2$ is at 
vanishing widths.
From these results, we suppose that 
the existence of sharp quasi-particle states in the quark spectral 
function is supported from the lattice analysis.
Fourth, the quark propagator depends on the gauge in general,
while the analyses on the lattice have been performed in Landau gauge.
We, however, note that the positions of poles in the quark propagator are 
independent of the gauge~\cite{Rebhan:1993az}.

\begin{figure}
\begin{center}
\includegraphics[width=0.49\textwidth]{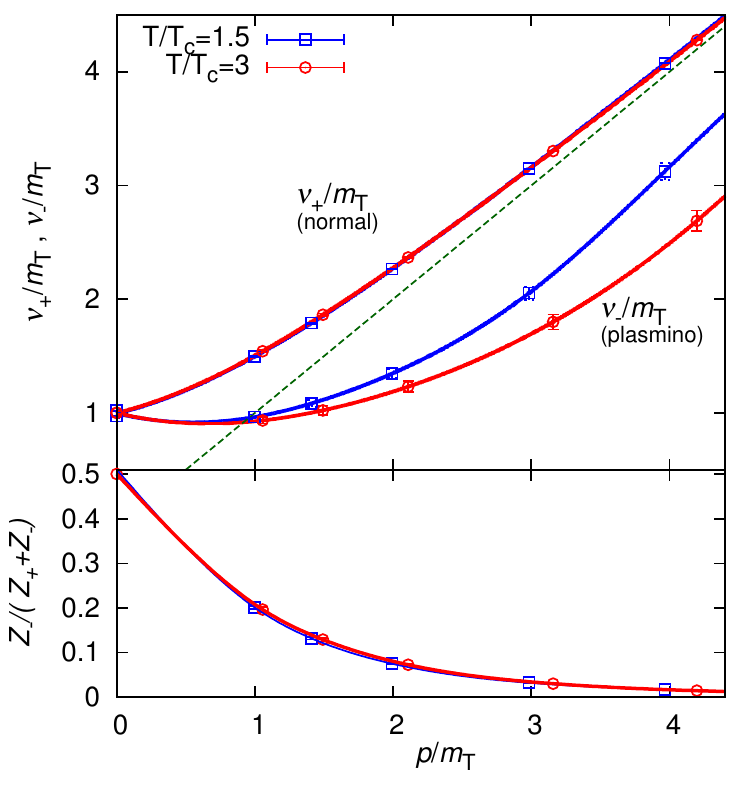}
\end{center}
\caption{
  Open symbols show the momentum dependence of the parameters 
  $\nu_+(p)$, $\nu_-(p)$, and $Z_-(p)/(Z_+(p)+Z_-(p))$ obtained 
  on the lattice in Ref.~\cite{Kaczmarek:2012mb}.
  The solid lines represent their interpolation obtained 
  by the cubic spline method.
  The dashed line in the upper panel represents the light cone.
}
\label{fig:LQPaSF_DISPERSION4}
\end{figure}

In Fig.~\ref{fig:LQPaSF_DISPERSION4},
we show the $p$ dependences of the fitting parameters in
Eq.~(\ref{eq:LQPaSF_2pole}) obtained in Ref.~\cite{Kaczmarek:2012mb}
for $T=1.5T_{\rm c}$ and $3T_{\rm c}$.
The upper panel shows the dispersions relation $\nu_+(p)$
and $\nu_-(p)$, while the lower panel shows
the relative strength of the residue of the plasmino mode
$Z_-(p)/(Z_+(p)+Z_-(p))$.
These analyses are performed on lattices with volume $128^3\times16$,
where both the lattice spacing and finite volume effects
are found to be small~\cite{Kaczmarek:2012mb}.
The two axes in the upper panel are normalized by the thermal mass $m_{\rm T}$
defined by the energy of the quasi-particle excitations at $p=0$ as 
$m_{\rm T}=(\nu_+(0)+\nu_-(0))/2$.
After the extrapolation to infinite volume,
$m_{\rm T}/T$ is estimated as 0.768(11) and 0.725(14)
at $T=1.5T_{\rm c}$ and $3T_{\rm c}$, respectively~\cite{Kaczmarek:2012mb}.

Although the lattice data are available only for discrete 
values of $p$ because of the periodic boundary condition, 
one needs the quark propagator as a 
continuous function of $p$ to solve Eq.~(\ref{eq:SDE_Pi_Start}).
To obtain the fitting parameters as continuous functions of $p$,
we perform the interpolation 
of the lattice data by the cubic spline method. 
From Eq.~(\ref{eq:rho_pm_mp}), one can show that 
$d\nu_+(p)/dp=-d\nu_-(p)/dp$, 
$d^2\nu_+(p)/dp^2=d^2\nu_-(p)/dp^2$, 
and $Z_+(p)=Z_-(p)$ at $p=0$ \cite{Karsch:2009tp,Kaczmarek:2012mb}.
These properties are taken into account in the interpolation.
The lattice data are available only in the momentum range 
$p/T \lesssim 4.7$, and we need 
the extrapolation to higher momentum.
We extrapolate the parameters using an exponentially damping form 
for $Z_-(p)/(Z_+(p)+Z_-(p))$, 
\begin{align}
  Z_-(p)/(Z_+(p)+Z_-(p)) = R{\rm e}^{-\alpha p},
  \label{eq:LQPaSF_extrapolate1}
\end{align}
and 
\begin{align}
  \nu_\pm(p) = p + \beta^\pm_1 {\rm e}^{-\beta^\pm_2 p},
  \label{eq:LQPaSF_extrapolate2}
\end{align}
for $\nu_\pm(p)$, which approaches the light cone exponentially for large $p$. 
The parameters $R$, $\alpha$, 
and $\beta^\pm_i$ are determined 
in the cubic spline analysis.
The $p$ dependence of each parameter determined in 
this way is shown by the solid lines in Fig.~\ref{fig:LQPaSF_DISPERSION4}.
We tested another extrapolation form, 
$\nu_\pm(p) = p + \beta'^\pm_1/p + \beta'^\pm_2/p^2 + \cdots$,
but found that the dispersion relations hardly change.
Finally, we set 
\begin{align}
  Z_+(p) + Z_-(p) = 1
\end{align}
to satisfy the sum rule Eq.~(\ref{eq:LQPaSF_SumRule}).

We note that the slope of the plasmino dispersion relation 
exceeds unity for $p \gtrsim 3 m_{\rm T}$ and is acausal.
This unphysical behavior may come from an artifact of the 
lattice simulation and/or the ansatz for the spectral function.
This acausal behavior gives rise to the Cherenkov radiation,
as we will see in later sections.
However, as we show explicitly in Sec.~\ref{sec:NR}, 
the residue of the plasmino mode is small, 
$Z_-(p)<0.05$, in this momentum range and 
the contribution of the Cherenkov radiation 
to our final result is negligibly small 
except for the low energy region $\omega\lesssim0.1$~GeV.

Using Eqs.~(\ref{eq:LQPaSF_S}), (\ref{eq:LQPaSF_Decompose}), and 
(\ref{eq:LQPaSF_2pole}), 
the quark propagator is given by 
\begin{align}
    S(i\nu_n,\bm{p}) &= S_+(i\nu_n,p) \Lambda_+ (\bm{p}) \gamma_0
  + S_-(i\nu_n,p) \Lambda_-(\bm{p}) \gamma_0,
  \nonumber \\
  &= \sum_{s=\pm1} S_s(i\nu_n,p) \Lambda_s (\bm{p}) \gamma_0 ,
  \label{eq:LQPaSF_S_lat}
\end{align}
with
\begin{align}
  S_s (i\nu_n,p)  =  \frac{ Z_+(p) }{ i\nu_n - s \nu_+(p) }
  + \frac{ Z_-(p) }{ i\nu_n + s \nu_-(p) },
  \label{eq:LQPaSF_S_s}
\end{align}
where we define $S_{\pm1}(P)=S_\pm(P)$ and
$\Lambda_{\pm1}(\bm{p})=\Lambda_\pm(\bm{p})$
for notational simplicity.
Correspondingly, the inverse quark propagator is given by
\begin{align}
  S^{-1}(i\nu_n,\bm{p})
  = \sum_{s=\pm1} S_s^{-1}(i\nu_n,p) \gamma_0  \Lambda_s(\bm{p}),
  \label{eq:LQPaSF_Sinv_lat}
\end{align}
with
\begin{align}
  S_s^{-1} (i\nu_n,p)
  =\frac{ (i\nu_n - s \nu_+(p))(i\nu_n + s \nu_-(p))}{i\nu_n + s V(p)},
  \label{eq:LQPaSF_partofSinv_lat}
\end{align}
and $V(p)=Z_+(p)\nu_-(p)-Z_-(p)\nu_+(p)$.
Note that the inverse propagator has poles at $i\nu_n=\pm V(p)$.
These poles inevitably appear in the multipole ansatz, because 
the propagator $S_s(\omega,p)$ after the analytic continuation to real 
time has one zero point in the range of $\omega$ surrounded by two poles.
We will see that the poles at $i\nu_n=\pm V(p)$ give rise to 
additional photon production mechanism 
that we call ``anomalous'' productions.

\subsection{Photon-Quark Vertex}
\label{sec:PQV}

To solve the SDE, Eq.~(\ref{eq:SDE_Pi_Start}), 
one needs the full photon-quark vertex $\Gamma_\mu(P+Q,P)$
besides the full quark propagator.
The vertex $\Gamma_\mu(P+Q,P)$ must satisfy the Ward-Takahashi 
identity (WTI)
\begin{align}
  Q^\mu\Gamma_\mu(P+Q,P)=S^{-1}(P+Q)-S^{-1}(P).
  \label{eq:PQV_WTI}
\end{align}
One, however, cannot determine the four components of the vertex
uniquely from this constraint.
In the present study, 
as the full vertex satisfying the WTI, we choose the following 
simple representation:
\begin{align}
  &\Gamma_0(i\omega_m+i\nu_n,\bm{p}+\bm{q};i\nu_n,\bm{p})
  \nonumber \\
  &= \frac{1}{2i\omega_m}\left[S^{-1}(i\omega_m+i\nu_n,\bm{p}+\bm{q})
    - S^{-1}(i\nu_n,\bm{p}+\bm{q})
    \right.
    \nonumber \\
    &~ ~ ~\left.
    + S^{-1}(i\omega_m+i\nu_n,\bm{p})
    - S^{-1}(i\nu_n,\bm{p})\right],
  \label{eq:PQV_G_0}
  \\
  &\Gamma_i(i\omega_m+i\nu_n,\bm{p}+\bm{q};i\nu_n,\bm{p})
  \nonumber \\
  &=\gamma_i
  -\frac{q_i}{2q^2}
  \left[
    S^{-1}(i\omega_m+i\nu_n,\bm{p}+\bm{q})
    + S^{-1}(i\nu_n,\bm{p}+\bm{q})
    \right.
    \nonumber \\
    &~ ~ ~\left.
    - S^{-1}(i\omega_m+i\nu_n,\bm{p})
    - S^{-1}(i\nu_n,\bm{p})
    \right]
  -\frac{q_i(\bm{q}\cdot\bm{\gamma})}{q^2}.
  \label{eq:PQV_Gamma_i}
\end{align}
One can easily check that Eqs.~(\ref{eq:PQV_G_0})
and (\ref{eq:PQV_Gamma_i}) satisfy the WTI Eq.~(\ref{eq:PQV_WTI}).
In this choice, we construct the vertex so that $\Gamma_\mu$ becomes
the bare vertex $\gamma_\mu$ when the free quark propagator 
$S^{-1}(P)=\slash{p}$ is substituted.
Although our vertex Eqs.~(\ref{eq:PQV_G_0}) and (\ref{eq:PQV_Gamma_i})
is not the unique choice \cite{Ball:1980ay}, we emphasize that 
our analysis with Eqs.~(\ref{eq:PQV_G_0}) and (\ref{eq:PQV_Gamma_i})
satisfies the gauge invariance required by the WTI.

\section{Photon Production Rate}
\label{sec:PPR}

Next, we construct the photon self-energy Eq.~(\ref{eq:SDE_Pi_Start})
from the lattice quark propagator and the gauge invariant
photon-quark vertex obtained in the previous section.
In this section,
we analyze the photon self-energy for two cases.
First, we calculate the photon production rate with the lattice quark 
propagator but with the bare vertex in Sec.~\ref{sec:PPRwVC}.
This result will help us understand the effect of the vertex correction.
Then, we perform the full analysis in Sec.~\ref{sec:FPPR}.

\subsection{Photon Production Rate without Vertex Correction}
\label{sec:PPRwVC}
\begin{figure}
\begin{center}
\includegraphics[width=0.33\textwidth]{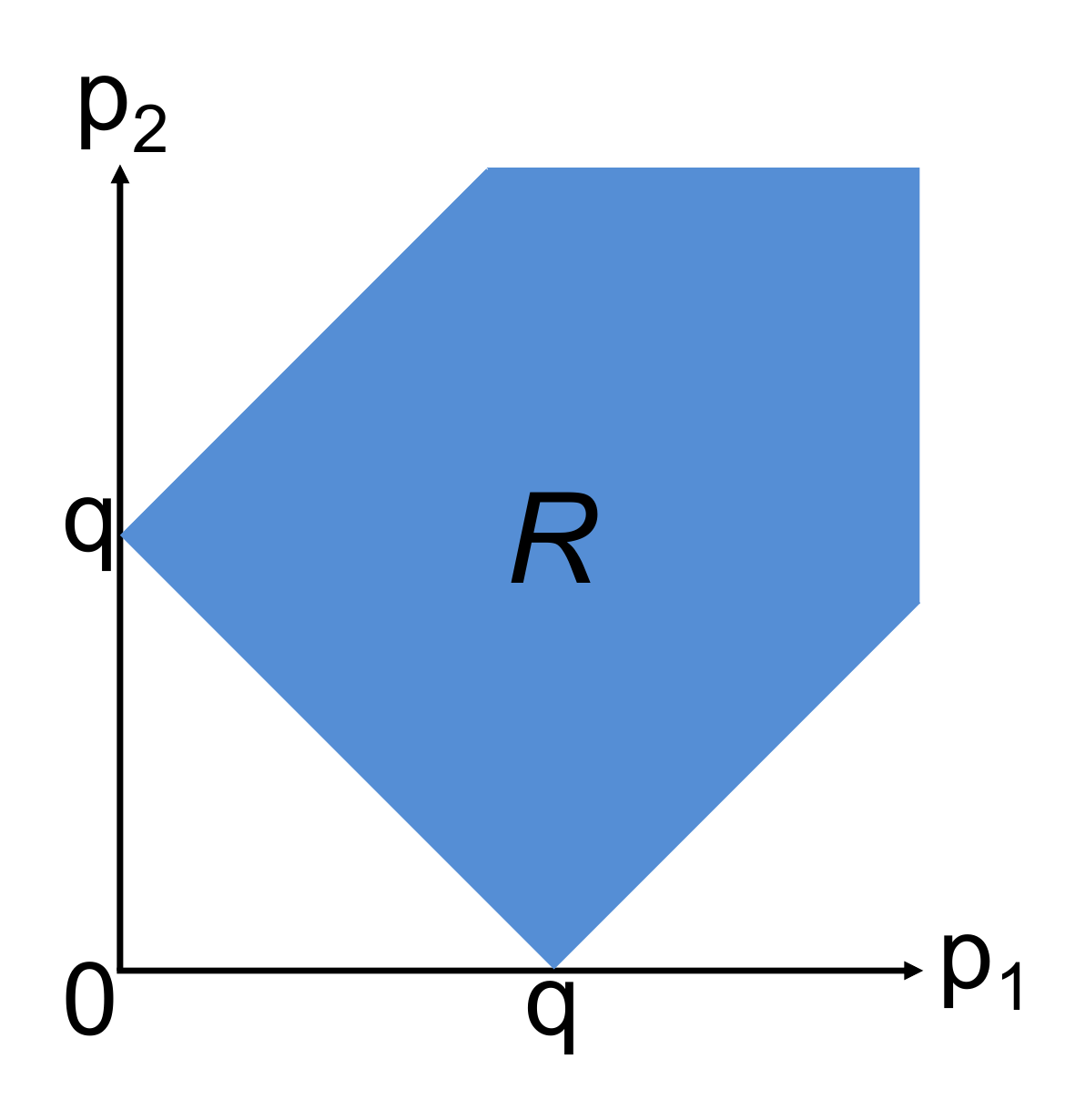}
\end{center}
\caption{
Integral region $R$ in Eq.~(\ref{eq:integral}).
}
\label{fig:PPRwVC_R}
\end{figure}

In this subsection,
we calculate the photon self-energy without the vertex correction 
with $\Gamma_\mu=\gamma_\mu$.
The photon self-energy in this case is given by
\begin{align}
  \Pi_{\mu\nu}(i\omega_m,\bm{q})=&
  -\frac{5e^2}{3}T\sum_n\int\frac{d^3p}{(2\pi)^3}
  \nonumber \\
  &{\rm Tr_D}[S(i\nu_n,\bm{p})\gamma_\mu S(i\omega_m+i\nu_n,\bm{p}+\bm{q})
    \gamma_\mu].
  \label{eq:PPRwVC_Pi_Start}
\end{align}
By substituting the decomposition Eq.~(\ref{eq:LQPaSF_S_lat}) into
Eq.~(\ref{eq:PPRwVC_Pi_Start}), one obtains
\begin{align}
  &\Pi_{\mu\mu}(i\omega_m,\bm{q})
  \nonumber \\
  &=-\frac{5e^2}{3}T\sum_n\int\frac{d^3p}{(2\pi)^3}
  \sum_{s,t=\pm1}S_s(i\nu_n,p)S_t(i\omega_m+i\nu_n,p)
  \nonumber \\
  & \qquad  {\rm Tr_D}[\Lambda_s(\bm{p})\gamma_0\gamma_\mu
  \Lambda_t(\bm{p}+\bm{q})\gamma_0\gamma_\mu]
  \label{eq:PPRwVC_Pi_TrD}
  \\ 
  &=-\frac{5e^2}{12\pi^2q}T\sum_n\sum_{s,t=\pm1}
  \int_R dp_1dp_2
  st[(sp_1-tp_2)^2-q^2]
  \nonumber \\
  & \qquad 
  S_s(i\nu_n,p_1)S_t(i\omega_m+i\nu_n,p_2),
  \label{eq:PPRwVC_Pi_mumu}
\end{align}
with 
\begin{align}
        p_1=|\bm{p}|, \quad p_2=|\bm{p}+\bm{q}|
\end{align}
being the momenta of the quarks, and $q=|\bm{q}|$, 
and we used
${\rm Tr_D}[\Lambda_s(\bm{p})\gamma_0\gamma_\mu
\Lambda_t(\bm{p}+\bm{q})\gamma_0\gamma_\mu] =
st[(sp_1-tp_2)^2-q^2]$.
In Eq.~(\ref{eq:PPRwVC_Pi_mumu}),
we changed the three-momentum integral as 
\begin{align}
  \int\frac{d^3p}{(2\pi)^3}=
  \int_R\frac{dp_1dp_2}{(2\pi)^2}
  \frac{p_1p_2}{q},
  \label{eq:integral}
\end{align}
where the integral is carried out in the region $R$,
which is surrounded by the lines $p_1+p_2=q$ and $p_1-p_2=\pm q$ as shown 
in Fig.~\ref{fig:PPRwVC_R}.
The region $R$ is defined by the $p_1$ and $p_2$ which can 
satisfy $|\bm{p}_2-\bm{p}_1|=|\bm{q}|$.

Taking the analytic continuation $i\omega_m\to\omega+i\eta$ of 
Eq.~(\ref{eq:PPRwVC_Pi_mumu}) and the imaginary part,
one obtains the imaginary part of the retarded self-energy,
\begin{align}
  {\rm Im}\Pi_\mu^{R,\mu}(\omega,\bm{q})
  =&\frac{5e^2}{12\pi q}\sum_{\substack{s,t,\eta_1,\eta_2\\=\pm1}}\int_R dp_1dp_2
  Z_{\eta_1}(p_1)Z_{\eta_2}(p_2)
  \nonumber \\
  &
  st[(sp_1-tp_2)^2-q^2]
  \nonumber \\
  &
  [f(s\eta_1\nu_{\eta_1}(p_1))-f(t\eta_2\nu_{\eta_2}(p_2))]
  \nonumber \\
  &
  \delta(\omega+s\eta_1\nu_{\eta_1}(p_1)-t\eta_2\nu_{\eta_2}(p_2)),
  \label{eq:PPRwVC_ImPi_mumu}
\end{align}
with $\nu_{\pm1}(p)=\nu_\pm(p)$ and
the Fermi distribution function $f(E)=1/[{\rm e}^{\beta E}+1]$.

\begin{figure}
\begin{center}
\includegraphics[width=0.49\textwidth]{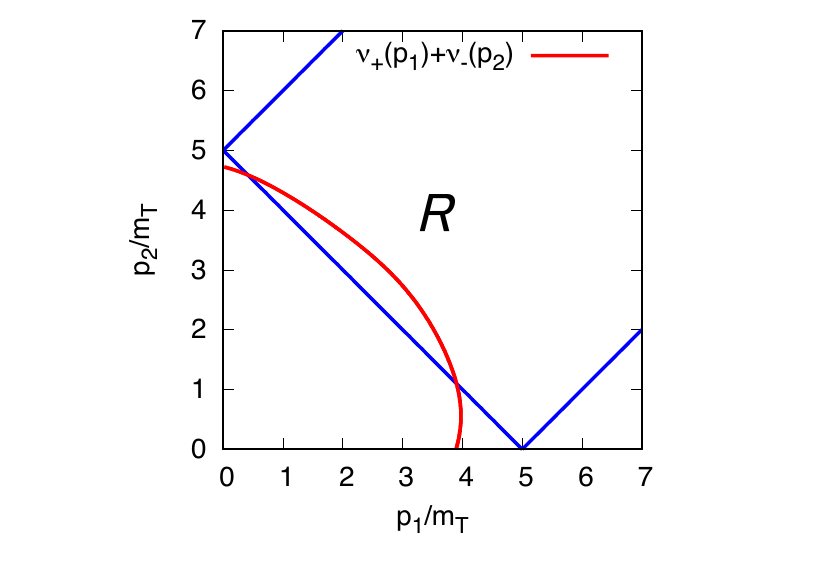}
\end{center}
\caption{
An example of the path of line integral 
in Eq.~(\ref{eq:PPRwVC_ImPi_mumu}) in the $p_1$--$p_2$ plane.
The red line shows the line constrained by the $\delta$-function 
for $\omega=|\bm{q}|=5m_{\rm T}$ and $s=t=-\eta_1=\eta_2=-1$ at $T=1.5T_{\rm c}$.
}
\label{fig:PPRwVC_np}
\end{figure}

Owing to the $\delta$-function, 
the momentum integral in Eq.~(\ref{eq:PPRwVC_ImPi_mumu}) is reduced
to the one-dimensional one.
In Fig.~\ref{fig:PPRwVC_np},
we show an example of the path determined by the $\delta$-function 
by the red line for $\omega=|\bm{q}|=5m_{\rm T}$ and 
$s=t=-\eta_1=\eta_2=-1$ at $T=1.5T_{\rm c}$; the line in this case
satisfies $\nu_+(p_1)+\nu_-(p_2)=5m_{\rm T}$.
In our numerical analysis, after determining the line 
in $R$ numerically we carry out the line integral. 

\begin{figure}
\begin{center}
\subfigure[Pair annihilation]{
\includegraphics*[width=0.22\textwidth]{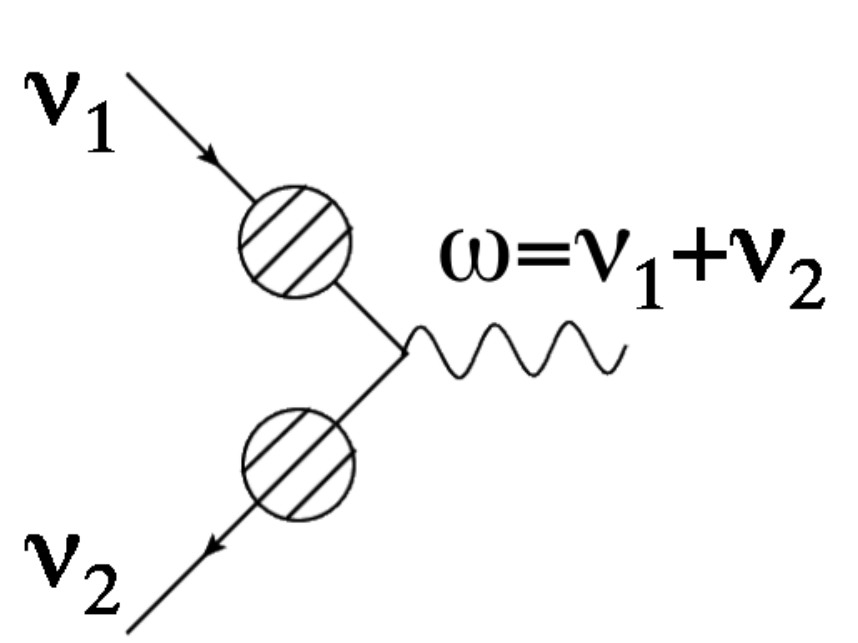}
\label{fig:PPRxVC_annihilation2}}
\subfigure[Landau damping]{
\includegraphics*[width=0.22\textwidth]{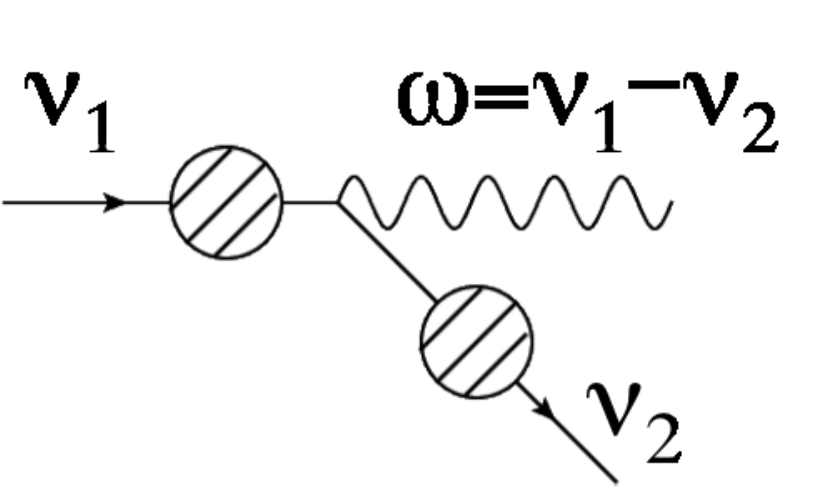}
\label{fig:PPRxVC_Landau2}}
\end{center}
\caption{
  Photon production processes with two quasi-quarks.
  Shaded circles represent the lattice quark propagator having two quasi-particle modes.
}
\label{fig:PPRxVC_decay}
\end{figure}

Equation~(\ref{eq:PPRwVC_ImPi_mumu}) consists of 16 terms with
$s,t,\eta_1,\eta_2=\pm1$.
These terms are interpreted as the photon production rates
from the decay and scatterings of two quasi-quarks as follows.
First, the $\delta$-function in each term 
expresses the energy conservation, and
the residues $Z_\pm$ represent those of the corresponding quasi-particles.
Next, the thermal factor
$f(s\eta_1\nu_{\eta_1}(p_1))-f(t\eta_2\nu_{\eta_2}(p_2))$
is transformed as 
$-[1-f(\nu_1)-f(\nu_2)]$ or $f(\nu_1)-f(\nu_2)$ for a 
given set of $(s,t,\eta_1,\eta_2)$
with $\nu_1=\nu_\pm(p_1)$ and $\nu_2=\nu_\pm(p_2)$.
When the thermal factor becomes
$-[1-f(\nu_1)-f(\nu_2)]$,
this factor can be rewritten as
\begin{align}
  &-[1-f(\nu_1)-f(\nu_2)]
  \nonumber \\
  &=f(\nu_1)f(\nu_2)-(1-f(\nu_1))(1-f(\nu_2)).
  \label{eq:PPRwVC_1-f-f}
\end{align}
Because this is the difference between the products of thermal 
distributions and Pauli blocking effects, 
this term is interpreted as the difference of the photon emission from
the pair annihilation of quasi-quarks 
shown in Fig.~\ref{fig:PPRxVC_annihilation2} and the photon absorption
mediated by its inverse process.
Similarly, we have 
\begin{align}
  f(\nu_1)-f(\nu_2)=f(\nu_1)(1-f(\nu_2))-(1-f(\nu_1))f(\nu_2).
  \label{eq:PPRwVC_f-f}
\end{align}
The photon emission and absorption of this term is diagrammatically
represented as in Fig.~\ref{fig:PPRxVC_Landau2},
which is called the Landau damping~\cite{Landau:1946jc}.

In the perturbative calculation, 
the pair annihilation with two quasi-particles cannot produce a real
photon~\cite{Kapusta:1991qp,Baier:1991em,Arnold:2001ba,Arnold:2001ms}.
This is because the quasi-particle states are in the time like
region and these processes cannot fulfill the on-shell condition
in the processes in Fig.~\ref{fig:PPRxVC_decay}.
On the other hand, as in Fig.~\ref{fig:LQPaSF_DISPERSION4} the plasmino
mode in the lattice quark propagator enters space-like region.
Because of this behavior, it is possible that the pair annihilation
produces a real photon directly.
This is the most remarkable difference of our analysis from the
perturbative one.

Note that we are interested in the photon production rate,
while ${\rm Im}\Pi^{R,\mu}_\mu$ represents the difference of the 
production and absorption rates as shown above.
To obtain the former, as in Eq.~(\ref{eq:SDE_PhotonRate}) 
one multiplies the Bose distribution function $1/[{\rm e}^{\beta\omega}-1]$
to ${\rm Im}\Pi_\mu^{R,\mu}(\omega,\bm{q})$. 
Indeed, we have
\begin{align}
  &\frac{1}{{\rm e}^{\beta\omega}-1}[1-f(\nu_1)-f(\nu_2)]
  \delta(\omega-\nu_1-\nu_2)
  \nonumber \\
  &=f(\nu_1)f(\nu_2)\delta(\omega-\nu_1-\nu_2).
\end{align}

\subsection{Full Photon Production Rate}
\label{sec:FPPR}

Now, we consider the photon self-energy
with the full vertex Eqs.~(\ref{eq:PQV_G_0}) and (\ref{eq:PQV_Gamma_i}).

First, we calculate ${\rm Im}\Pi_{00}(\omega,\bm{q})$.
The trace in Eq.~(\ref{eq:SDE_Pi_Start}) is calculated to be
\begin{widetext}
\begin{align}
  &{\rm Tr_D}[S(P_1)\gamma_0S(P_2)\Gamma_0(P_2,P_1)]
  \nonumber \\
  &=\sum_{s,t=\pm1}\frac{1}{2i\omega_m}{\rm Tr_D}[\Lambda_s(p_1)\Lambda_t(p_2)]
  \left[
    S_s(i\nu_n,p_1)
    \left[
    1-\frac{S_t(i\omega_m+i\nu_n,p_2)}{S_t(i\nu_n,p_2)}
    \right]
    -S_t(i\omega_m+i\nu_n,p_2)
    \left[
    1-\frac{S_s(i\nu_n,p_1)}{S_s(i\omega_m+i\nu_n,p_1)}
    \right]
    \right]
  \nonumber \\
  &=\sum_{s,t,\epsilon=\pm1}\frac{1}{2}{\rm Tr_D}[\Lambda_s(p_1)\Lambda_t(p_2)]
  \left[  
    S_s(i\nu_n,p_1)
    \frac{Z_{\epsilon}(p_2)(i\nu_n+t\epsilon\nu_{\bar{\epsilon}}(p_2))}{(i\omega_m+i\nu_n-\epsilon t\nu_{\epsilon}(p_2))(i\nu_n+tV(p_2))}
    \right.
    \nonumber \\
    &~ ~ ~ ~
    \left.
    +
    S_t(i\omega_m+i\nu_n,p_2)
    \frac{Z_{\epsilon}(p_1)(i\omega_m+i\nu_n+s\epsilon\nu_{\bar{\epsilon}}(p_1))}{(i\nu_n-\epsilon s\nu_{\epsilon}(p_1))(i\omega_m+i\nu_n+sV(p_1))}
    \right]
  \label{eq:FPPR_Sg0SG0}
  ,
\end{align}
with $\nu_{\bar{\epsilon}}=\nu_{-\epsilon}$, 
where in the first and second equalities we used, respectively,
the decomposition Eq.~(\ref{eq:LQPaSF_S_lat}) and the two-pole form 
Eq.~(\ref{eq:LQPaSF_S_s}).
Using $V(p)+\eta\nu_\eta(p)=\eta Z_\eta(p)\bar{\nu}(p)$ and
$V(p)+\bar{\eta}\nu_{\bar{\eta}}(p)=\bar{\eta} Z_{\bar{\eta}}(p)\bar{\nu}(p)$
with $\bar{\nu}(p)=\nu_+(p)+\nu_-(p)$, 
$\Pi_{00}(i\omega_m,\bm{q})$ is given by
\begin{align}
  &\Pi_{00}(i\omega_m,\bm{q})
  =-\frac{5e^2}{12(2\pi)^2q}\sum_{\substack{s,t,\eta_1,\eta_2\\=\pm1}}\int_R dp_1dp_2
  st[(sp_1+tp_2)^2-q^2]Z_{\eta_1}(p_1)Z_{\eta_2}(p_2)
  \nonumber \\
  &~ ~ ~ ~
  \left[\left(
    2+\frac{t\eta_2Z_{\bar{\eta}_2}(p_2)\bar{\nu}(p_2)}{s\eta_1\nu_{\eta_1}(p_1)+tV(p_2)}
    +\frac{s\eta_1Z_{\bar{\eta}_1}(p_1)\bar{\nu}(p_1)}{t\eta_2\nu_{\eta_2}(p_2)+sV(p_1)}
    \right)\right.
    \frac{f(s\eta_1\nu_{\eta_1}(p_1))-f(t\eta_2\nu_{\eta_2}(p_2))}{i\omega_m+s\eta_1\nu_{\eta_1}(p_1)-t\eta_2\nu_{\eta_2}(p_2)}
  \nonumber \\
  &~ ~ ~ ~
  \left.      
  -\frac{t\eta_2Z_{\bar{\eta}_2}(p_2)\bar{\nu}(p_2)}{s\eta_1\nu_{\eta_1}(p_1)+tV(p_2)}
  \cdot\frac{f(-tV(p_2))-f(t\eta_2\nu_{\eta_2}(p_2))}{i\omega_m-tV(p_2)-t\eta_2\nu_{\eta_2}(p_2)}
  -\frac{s\eta_1Z_{\bar{\eta}_1}(p_1)\bar{\nu}(p_1)}{t\eta_2\nu_{\eta_2}(p_2)+sV(p_1)}
  \cdot\frac{f(s\eta_1\nu_{\eta_1}(p_1))-f(-sV(p_1))}{i\omega_m+s\eta_1\nu_{\eta_1}(p_1)+sV(p_1)}
  \right].
  \label{eq:FPPR_Pi_00}
\end{align}
Performing the analytic continuation and taking the imaginary part,
one obtains ${\rm Im}\Pi_{00}^R(\omega,\bm{q})$ as
\begin{align}
  &{\rm Im}\Pi_{00}^R(\omega,\bm{q})
  =\frac{5\alpha}{12q}\sum_{\substack{s,t,\eta_1,\eta_2\\=\pm1}}\int_R dp_1dp_2
  st[(sp_1+tp_2)^2-q^2]Z_{\eta_1}(p_1)Z_{\eta_2}(p_2)
  \nonumber \\
  &~ ~ ~ ~
  \left[\left(
    2+\frac{t\eta_2Z_{\bar{\eta}_2}(p_2)\bar{\nu}(p_2)}{s\eta_1\nu_{\eta_1}(p_1)+tV(p_2)}
    +\frac{s\eta_1Z_{\bar{\eta}_1}(p_1)\bar{\nu}(p_1)}{t\eta_2\nu_{\eta_2}(p_2)+sV(p_1)}
    \right)\right.
  [f(s\eta_1\nu_{\eta_1}(p_1))-f(t\eta_2\nu_{\eta_2}(p_2))]
  \delta(\omega+s\eta_1\nu_{\eta_1}(p_1)-t\eta_2\nu_{\eta_2}(p_2))
  \nonumber \\
  &~ ~ ~ ~
  -\frac{t\eta_2Z_{\bar{\eta}_2}(p_2)\bar{\nu}(p_2)}{s\eta_1\nu_{\eta_1}(p_1)+tV(p_2)}
  [f(-tV(p_2))-f(t\eta_2\nu_{\eta_2}(p_2))]
  \delta(\omega-tV(p_2)-t\eta_2\nu_{\eta_2}(p_2))
  \nonumber \\
  &~ ~ ~ ~
  \left.
  -\frac{s\eta_1Z_{\bar{\eta}_1}(p_1)\bar{\nu}(p_1)}{t\eta_2\nu_{\eta_2}(p_2)+sV(p_1)}
  [f(s\eta_1\nu_{\eta_1}(p_1))-f(-sV(p_1))]
  \delta(\omega+s\eta_1\nu_{\eta_1}(p_1)+sV(p_1))
  \right].
  \label{eq:FPPR_ImPi00}
\end{align}

\end{widetext}

Next, we analyze $\sum_i{\rm Im}\Pi_{ii}^R$.
We decompose $\sum_i \Pi_{ii}$ into three parts,
$\Pi_{ii}(Q)=\sum_{j=1}^3\Pi_{ii}^{(j)}(Q)$, where
\begin{align}
  \Pi_{ii}^{(j)}(Q)=-\frac{5}{3}T\sum_n\int\frac{d^3p}{(2\pi)^3}
  {\rm Tr_D}[S(P)\gamma_iS(P+Q)\Gamma_i^{(j)}],
  \label{eq:Pi_ii^(j)}
\end{align}
with
\begin{align}
  \Gamma_i^{(1)}&=\gamma_i,
  \\
  \Gamma_i^{(2)}&=-\frac{q_i}{2q^2}
  \left[
    S^{-1}(i\omega_m+i\nu_n,\bm{p}+\bm{q})
    + S^{-1}(i\nu_n,\bm{p}+\bm{q})
    \right.
    \nonumber
    \\
    &~ ~ ~ \left.
    - S^{-1}(i\omega_m+i\nu_n,\bm{p})
    - S^{-1}(i\nu_n,\bm{p})
    \right],
  \\
  \Gamma_i^{(3)}&=
  -\frac{q_i(\bm{q}\cdot\bm{\gamma})}{q^2}.
\end{align}  

The analysis of $\sum_i{\rm Im}\Pi_{ii}^{(1),R}(\omega,\bm{q})$
can be carried out in the same way as in Sec.~\ref{sec:PPRwVC}.
The result is given by
\begin{widetext}
\begin{align}
  \sum_{i=1}^3{\rm Im}\Pi_{ii}^{(1),R}(\omega,\bm{q})
  =&-\frac{5\alpha}{3q}\sum_{\substack{s,t,\eta_1,\eta_2\\=\pm1}}\int_R dp_1dp_2
  Z_{\eta_1}(p_1)Z_{\eta_2}(p_2)
  \left[\frac{st[(sp_1+tp_2)^2-q^2]}{2}-4p_1p_2\right]
  \nonumber \\
  &
  [f(s\eta_1\nu_{\eta_1}(p_1))-f(t\eta_2\nu_{\eta_2}(p_2))]
  \delta(\omega+s\eta_1\nu_{\eta_1}(p_1)-t\eta_2\nu_{\eta_2}(p_2)).
  \label{eq:FPPR_ImPiiiBare}
\end{align}

Next, we consider $\sum_i{\rm Im}\Pi_{ii}^{(2),R}(\omega,\bm{q})$.
The trace of this term is calculated to be
\begin{align}
  &{\rm Tr_D}[S(P)\gamma_iS(P+Q)
  [S^{-1}(i\omega_m+i\nu_n,\bm{p}+\bm{q})
  +S^{-1}(i\omega_m,\bm{p}+\bm{q})
  -S^{-1}(i\omega_m+i\nu_n,\bm{p})
  -S^{-1}(i\nu_n,\bm{p})]]
  \nonumber \\
  &=\sum_{s,t}{\rm Tr_D}[\Lambda_s(\bm{p})\gamma_0\gamma_i\Lambda_t(\bm{p}+\bm{q})]
  \left[
    S_s(i\nu_n,p_1)
    \left[
      1+\frac{S_t(i\omega_m+i\nu_n,p_2)}{S_t(i\nu_n,p_2)}
      \right]
    -S_t(i\omega_m+i\nu_n,p_2)
    \left[
      1+\frac{S_s(i\nu_n,p_1)}{S_s(i\omega_m+i\nu_n,p_1)}
      \right]
  \right]
  \nonumber \\
  &\to\sum_{s,t}{\rm Tr_D}[\Lambda_s(\bm{p})\gamma_0\gamma_i\Lambda_t(\bm{p}+\bm{q})]
  \left[
    S_s(i\nu_n,p_1)
    \left[
      \frac{S_t(i\omega_m+i\nu_n,p_2)}{S_t(i\nu_n,p_2)}-1
      \right]
    -S_t(i\omega_m+i\nu_n,p_2)
    \left[
      \frac{S_s(i\nu_n,p_1)}{S_s(i\omega_m+i\nu_n,p_1)}-1
      \right]
  \right],
  \label{eq:Tr_ii}
\end{align}
where in the final step we used the fact that 
$S_s(i\nu_n,p_1)-S_t(i\omega_m+i\nu_n,p_2)$ vanishes after this term
is integrated out and summed over $s$ and $t$ in Eq.~(\ref{eq:Pi_ii^(j)}).
Using Eq.~(\ref{eq:Tr_ii}) one finds that 
$\sum_i{\rm Im}\Pi_{ii}^{(2),R}(\omega,\bm{q})$ has the same form
as in Eq.~(\ref{eq:FPPR_ImPi00}) but the factor 
$\omega(tp_2-sp_1)/q^2$ is applied to the integrand.

The last piece $\sum_i{\rm Im}\Pi_{ii}^{(3),R}(\omega,\bm{q})$
is calculated as
\begin{align}
  \sum_i{\rm Im}\Pi_{ii}^{(3),R}
  =&
  \frac{5\alpha}{3q}\sum_{\substack{s,t,\eta_1,\eta_2\\=\pm1}}\int_R dp_1dp_2
  p_1p_2Z_{\eta_1}(p_1)Z_{\eta_2}(p_2)
  \left[1-\frac{st}{2p_1p_2q^2}\left[(p_1^2+p_2^2)q^2
    -(p_1^2-p_2^2)^2\right]\right]
  \nonumber \\
  &\times
  [f(s\eta_1\nu_{\eta_1}(p_1))-f(t\eta_2\nu_{\eta_2}(p_2))]
  \delta(\omega+s\eta_1\nu_{\eta_1}(p_1)-t\eta_2\nu_{\eta_2}(p_2)).
  \label{eq:FPPR_ImPi(3)ii}
\end{align}

Finally, by taking the trace of ${\rm Im}\Pi_{\mu\nu}^R(\omega,\bm{q})$,
we obtain the full photon self-energy as
\begin{align}
  &{\rm Im}\Pi_\mu^{R,\mu}(\omega,\bm{q})
  =\frac{5\alpha}{12q}\sum_{\substack{s,t,\eta_1,\eta_2\\=\pm1}}\int_R dp_1dp_2
  stZ_{\eta_1}(p_1)Z_{\eta_2}(p_2)
  \biggl\{
    [(sp_1+tp_2)^2-q^2]
    \left(1-\frac{\omega(tp_2-sp_1)}{q^2}\right)
    \nonumber \\
    &\times
    \left[
      -\frac{t\eta_2Z_{\bar{\eta}_2}(p_2)\bar{\nu}(p_2)}{s\eta_1\nu_{\eta_1}(p_1)+tV(p_2)}
      [f(-tV(p_2))-f(t\eta_2\nu_{\eta_2}(p_2))]
      \delta(\omega-tV(p_2)-t\eta_2\nu_{\eta_2}(p_2))
      \right.
      \nonumber \\
      &\hspace{10mm}
      \left.
      -\frac{s\eta_1Z_{\bar{\eta}_1}(p_1)\bar{\nu}(p_1)}{t\eta_2\nu_{\eta_2}(p_2)+sV(p_1)}
      [f(s\eta_1\nu_{\eta_1}(p_1))-f(-sV(p_1))]
      \delta(\omega+s\eta_1\nu_{\eta_1}(p_1)+sV(p_1))
      \right]
    \nonumber \\
   &+\biggl[[(sp_1+tp_2)^2-q^2]
      \left(1-\frac{\omega(tp_2-sp_1)}{q^2}\right)
      \left(2
      +\frac{t\eta_2Z_{\bar{\eta}_2}(p_2)\bar{\nu}(p_2)}{s\eta_1\nu_{\eta_1}(p_1)+tV(p_2)}
      +\frac{s\eta_1Z_{\bar{\eta}_1}(p_1)\bar{\nu}(p_1)}{t\eta_2\nu_{\eta_2}(p_2)+sV(p_1)}
      \right)
      \nonumber \\
      &
      +\frac{2}{q^2}[(sp_1+tp_2)^2+q^2][(sp_1-tp_2)^2-q^2]
      \biggl]
      [f(s\eta_1\nu_{\eta_1}(p_1))-f(t\eta_2\nu_{\eta_2}(p_2))]
      \delta(\omega+s\eta_1\nu_{\eta_1}(p_1)-t\eta_2\nu_{\eta_2}(p_2))
    \biggl\}.
  \label{eq:FPPR_ImPiVC}
\end{align}
\end{widetext}
The photon production rate is obtained by substituting 
this result into Eq.~(\ref{eq:SDE_PhotonRate}).

Let us inspect the physical meaning of each term in Eq.~(\ref{eq:FPPR_ImPiVC}).
Similar to the discussion in Sec.~\ref{sec:PPRwVC},
the terms in the last two lines in Eq.~(\ref{eq:FPPR_ImPiVC})
are identified to be the pair annihilation and Landau damping of 
quasi-particles from the structures of the $\delta$-function and thermal factor.
These terms have different coefficients from those in 
Eq.~(\ref{eq:PPRwVC_ImPi_mumu}) owing to the vertex correction.
The terms in the first three lines in Eq.~(\ref{eq:FPPR_ImPiVC})
seem to be photon emissions mediated by the 
quasi-particles having energies $\nu_\eta$ and $V$.
However, the lattice quark propagator 
does not have the quasi-particle mode with the dispersion~$\nu=V(p)$.
Therefore, the photon productions expressed in these terms 
cannot be interpreted as simple reactions
between quasi-particles and a photon.
These photon productions come from the anomalous pole included 
in the inverse propagator $S_s^{-1}$ in the vertex function 
Eqs.~(\ref{eq:PQV_G_0}) and (\ref{eq:PQV_Gamma_i}) \cite{Kim:2015poa}.
Although these terms are peculiar at a glance, they naturally appear
to satisfy the gauge invariance and thus must be included in the 
full photon production rate.

\section{Numerical results}
\label{sec:NR}

\begin{figure}
\begin{center}
\includegraphics[width=0.49\textwidth]{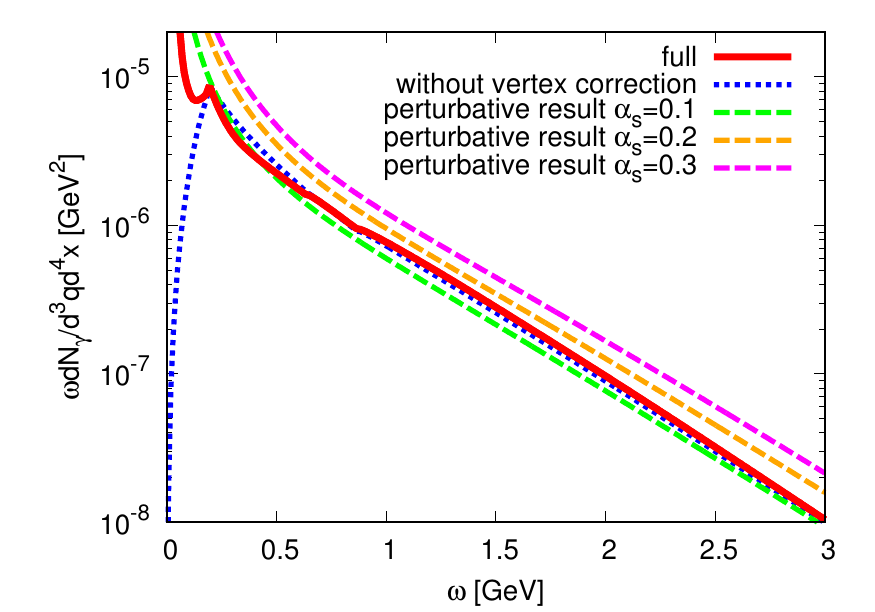}
\end{center}
\caption{
Photon production rate for $T=1.5T_{\rm c}$.
The result without vertex correction is expressed by the dotted line
and perturbative results are plotted with the dashed lines.
}
\label{fig:NR_1ten5}
\end{figure}

Next, let us see the numerical results on the
photon production spectrum.
In Fig.~\ref{fig:NR_1ten5},
we show the photon production rate with the lattice quark propagator and 
the gauge invariant vertex as a function of the photon energy~$\omega$
for $T=1.5T_{\rm c}$ by the solid line.
The production rate with the bare vertex obtained
in Sec.~\ref{sec:PPRwVC} is also plotted by the dotted line.
The photon production rate has been calculated in the perturbative 
analysis~\cite{Arnold:2001ba,Arnold:2001ms,Ghiglieri:2013gia}.
In the figure, we show fitting formula for the perturbative result 
in Refs.~\cite{Arnold:2001ba,Arnold:2001ms} for three values 
of the strong coupling constant $\alpha_s=0.1,0.2$, and $0.3$
by the dashed lines;
this fitting function well reproduces the complete leading log result
for $0.2\leq \omega/T \leq 50$~\cite{Arnold:2001ms}.
The next-to-leading order result~\cite{Ghiglieri:2013gia}
enhances the rate at most 20\% around $\omega /T\sim 2$.

From Fig.~\ref{fig:NR_1ten5}, one finds that the production rates 
with full and bare vertices behave similarly for $\omega\gtrsim0.2$~GeV.
This result suggests that the effect of the vertex correction is 
not large.
The figure also shows that the production rate with the lattice
quark propagator behaves quite similarly with the perturbative result
with $\alpha_{\rm s} \simeq 0.1-0.2$
except for low energy region.
This result is unexpected, because the production mechanisms of photons
are different between these analyses.

\begin{figure}
\begin{center}
\includegraphics[width=0.49\textwidth]{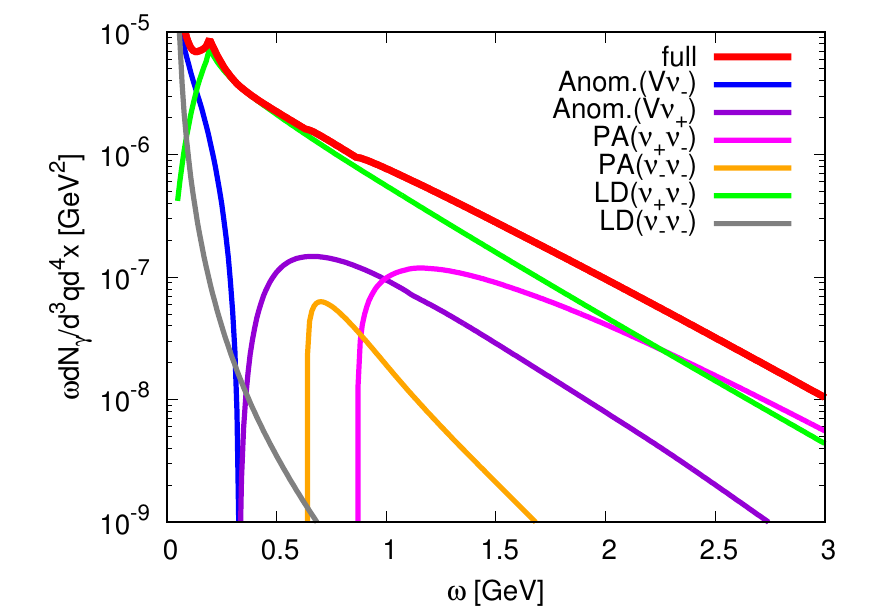}
\end{center}
\caption{
Decomposition of the photon production rate for $T=1.5T_{\rm c}$.
}
\label{fig:NR_Decomposed_1ten5}
\end{figure}

As discussed in the previous section, the photon production rate 
can be decomposed into contributions of different processes.
In Fig.~\ref{fig:NR_Decomposed_1ten5},
we show the contributions of each term in Eq.~(\ref{eq:FPPR_ImPiVC})
on the production rate.
In the figure, the lines show the rate 
from the pair annihilation of a normal and a plasmino
modes~(PA($\nu_+\nu_-$)) and two plasmino modes~(PA($\nu_-\nu_-$)),
the Landau damping between a normal and a plasmino
modes~(LD($\nu_+\nu_-$)) and two plasmino modes~(LD($\nu_-\nu_-$)),
the rates of two anomalous 
productions~(Anom.($V\nu_+$) and Anom.($V\nu_-$)), and
the total (full) rate.
The figure shows that the photon production for $\omega\lesssim 1.5$~GeV
is dominated by the Landau damping LD($\nu_+\nu_-$),
while for higher energies two production mechanisms 
LD($\nu_+\nu_-$) and PA($\nu_+\nu_-$) have almost the same contributions.
The process LD($\nu_-\nu_-$) is interpreted as the Cherenkov radiation
from the acausal dispersion relation of the plasmino mode.
From Fig.~\ref{fig:NR_Decomposed_1ten5}, one finds that the 
photon production with this process is well suppressed 
except for $\omega\lesssim0.2$~GeV.
This result shows that the effect of this unphysical process 
is negligible.
The figure also shows that the contribution of the anomalous productions 
(Anom.($V\nu_+$) and Anom.($V\nu_-$)) are 
small compared to other processes except for the low energy region
$\omega\lesssim0.2$~GeV.
The decomposition of the photon production rate obtained
with the bare vertex, Eq.~(\ref{eq:PPRwVC_ImPi_mumu}),
shows the similar tendency as the one of the full result.

\begin{figure}
\begin{center}
\includegraphics[width=0.49\textwidth]{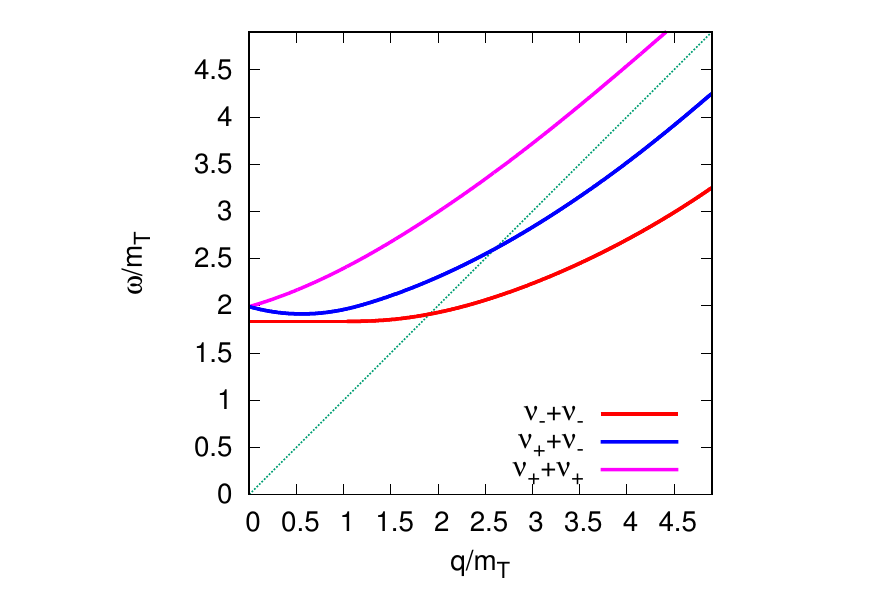}
\end{center}
\caption{
Dispersions of photons produced via quasi-particle pair annihilations.
Each line shows the minimum value of dispersions.
Dotted line is the light cone.
}
\label{fig:NR_PairDispersion}
\end{figure}

Figure~\ref{fig:NR_Decomposed_1ten5} shows 
that the processes PA$(\nu_+\nu_-)$ and PA$(\nu_-\nu_-)$
take nonzero values for $\omega>\omega_{+-}$ and $\omega_{--}$,
respectively, with $\omega_{+-}\simeq0.86$~GeV and 
$\omega_{--}\simeq0.63$~GeV.
Owing to the emergence of these processes, 
the total production rate has non-smooth behaviors at these energies.
To understand the reason why these processes 
emerge at $\omega=\omega_{+-}$ and $\omega_{--}$, 
in Fig.~\ref{fig:NR_PairDispersion} we show
the minimum of the photon energy produced by these processes
with a fixed photon momentum $q$.
In the figure, the minimum energy for the pair annihilation of two 
normal modes ($\nu_+\nu_+$) is also plotted for comparison.
In the regions above the lines, 
virtual photon production is allowed kinematically.
From the figure, one finds that the region covers even a part of the
space-like region for $\nu_+\nu_-$ and $\nu_-\nu_-$, 
and thus these processes can produce a real photon, while 
the region for $\nu_+\nu_+$ does not have an overlap with the light cone.
The real photon production from two quasi-particle pair 
annihilation does not occur in the perturbative calculation 
based on the hard thermal loop resummation technique,
because the normal and plasmino modes in this formalism 
exists only in the time-like region.
The lattice propagator, on the other hand, has the plasmino mode 
which enters space-like region as in Fig.~\ref{fig:LQPaSF_DISPERSION4}. 
This behavior is responsible for the emergence of the nonzero production 
rates by PA$(\nu_+\nu_-)$ and PA$(\nu_-\nu_-)$.

The photon spectra shown in Fig.~\ref{fig:NR_1ten5} has a cusp structure 
at $\omega=\omega_{\rm max}$ with $\omega_{\rm max}\simeq0.2$~GeV.
According to Fig.~\ref{fig:NR_Decomposed_1ten5}, this cusp comes from 
the Landau damping LD($\nu_+\nu_-$).
The origin of this structure is understood from the kinematics 
between quasi-particle pair and photon.
Let us denote the momenta of the normal and plasmino modes 
$\bm{p}_+$ and $\bm{p}_-$, respectively, and the photon
momentum $\bm{q}$. Then, these momenta satisfy 
$\bm{p}_+= \bm{p}_-+\bm{q}$.
For $\omega<\omega_{\rm max}$, the decay process is allowed for all
directions of quark momenta. At $\omega=\omega_{\rm max}$, 
the process with perpendicular $\bm{p}_+$ and $\bm{q}$, i.e.
the case with $|\bm{p}_+| + |\bm{q}| = |\bm{p}_-|$ starts to 
be forbidden kinematically, and the phase space becomes narrower
as $\omega=|\bm{q}|$ becomes larger.
This constraint in the phase space gives rise to the cusp in 
the production spectrum.

It is a surprising result that the photon production spectrum 
in our analysis has a similar behavior, not only in the slope but 
also the absolute value, 
as the perturbative results for almost all energy range.
The slopes in both results are approximately given by the 
Boltzmann factor ${\rm e}^{-\beta\omega}$.
The origin of this slope in our analysis can be understood as follows.
From Fig.~\ref{fig:NR_Decomposed_1ten5}, the photon production is 
dominated by two processes, PA($\nu_+\nu_-$) and LD($\nu_+\nu_-$).
These processes include the plasmino mode whose
residue~$Z_-$ is exponentially damping for large momentum.
Because of this factor, the momentum of the plasmino mode must be 
small, $\bm{p}_- \simeq0$.
Then, to produce a real photon with large momentum
$|\bm{q}|\gg m_{\rm T}$, 
the momentum of the normal mode must satisfy 
$\bm{p}_+=\bm{q}\pm\bm{p}_-\simeq\bm{q}$,
and the photon emission takes place via the decay of this mode.
The rate of this process is proportional to the thermal probability 
of the existence of the normal mode given by 
${\rm e}^{-\beta \nu_+(q)}\simeq{\rm e}^{-\beta q}={\rm e}^{-\beta \omega}$.
For sufficiently large $\bm{q}$, this factor determines the production
rate, because other effects, such as the phase space determined by 
kinematics and $Z_+(|\bm{p}_+|)\simeq1$, are insensitive to $\bm{q}$.
In the perturbative analysis, 
the photon self-energy at high energy is dominated by two-loop diagrams
and collinear enhanced diagrams \cite{Arnold:2001ms}.
In these processes, an initial quark with large momentum 
$\bm{q}$ must decay into soft degrees of freedom, and the 
Boltzmann factor appears for the same reason as ours, while 
it is known that other factors are insensitive to $\bm{q}$ 
\cite{Arnold:2001ms}. 
To summarize, the coincidence in the slope comes from the 
thermal factor for the initial quark with large momentum.
Because the decay processes in these analyses are completely different,
however, the coincidence in magnitude would be 
accidental.

\begin{figure}
\begin{center}
\includegraphics[width=0.49\textwidth]{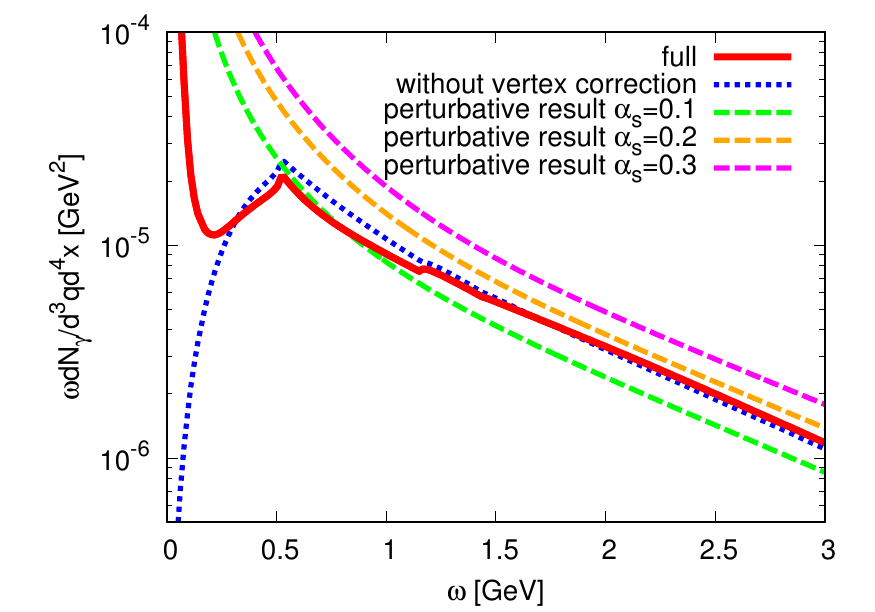}
\end{center}
\caption{
Photon production rate for $T=3T_{\rm c}$.
The result without vertex correction is expressed by the dotted line.
Perturbative results are plotted with the dashed lines for three values
of $\alpha_s$.
}
\label{fig:NR_3}
\end{figure}

In Fig.~\ref{fig:NR_3},
we show the result for $T=3T_{\rm c}$.
Its behavior is qualitatively the same as the production rate 
for $T=1.5T_{\rm c}$, although the spectrum has a clearer dip structure
at low energy $\omega\simeq0.5$~GeV.
The contributions of different decay processes behave similarly, too.

The results at both temperatures have a similar
form and magnitude with the perturbative results. If a closer
look is taken, however, we recognize that our non-perturbative results are
almost reproduced by perturbative results with
$\alpha_{\rm s} \simeq 0.1 - 0.2$, which is rather on the lower
side. This has an important phenomenological consequence;
the discrepancy between the observed photon yield at RHIC
and the corresponding theoretical result is not resolved, but
is enlarged. Thus, it is confirmed also by this investigation
that there will exist still unknown mechanisms for photon production
in relativistic heavy ion collisions.

\section{Summary}
\label{sec:Summary}

In this paper, 
we analyzed the production spectrum of thermal photons 
from the deconfined medium near the critical temperature $T_{\rm c}$.
In order to include the non-perturbative effect in the analysis, 
we analyzed the Schwinger-Dyson equation of the
photon self-energy with the quark propagator obtained
on the lattice for two temperatures.
The photon-quark vertex is constructed so as to satisfy the WTI.
The effect of the vertex correction is discussed.
We showed that the vertex function satisfying the WTI generates 
anomalous photon production mechanisms 
but its contribution is small.
In comparison with the perturbative analysis,
the spectra with the lattice quark propagator show almost the same 
behavior for $\omega\gtrsim0.5$~GeV.
We discussed that the coincidence in the slope can be 
understood from the production mechanism.
However, it is a nontrivial result that the our result is similar to 
the perturbative one even in the magnitude.
The gap between experimental data and theoretical calculations
on the yield and flows of photons still remain. Our result
indicates the existence of unknown mechanisms of photon production
in relativistic heavy ion collisions.

The authors thank T.~Shimoda, K.~Ogata, and K.~Oda for encouragements.
T.~K. is supported by Grant-in-Aid for JSPS research
fellows~(No.~16J00981) and
Osaka University Cross-Boundary Innovation Program.
This work is supported in part by JSPS KAKENHI Grant
Numbers 26400272 and 17K05442.

\end{document}